\documentclass[prd,showpacs,preprint]{revtex4}

\begin{document}

\title{Cosmological Constant Influence On Cosmic String Spacetime}

\author{Amir H. Abbassi}
\email{ahabbasi@modares.ac.ir}
\affiliation{Department of Physics, School of Sciences, Tarbiat Modarres
University, P.O.Box 14155-4838, Tehran, Iran.}
\author{Amir M. Abbassi}
\email{amabasi@khayam.ut.ac.ir}
\affiliation{Depatment of Physics, Faculty of Sciences, Tehran University, Tehran, Iran.}
\author{H. Razmi }
\email{razmi@qom.ac.ir}
\affiliation{Department of Physics, Faculty of Sciences, Qom University,
Qom, Iran.\\
	 {\rm (Received .....)}}

\begin{abstract}
 We are going to investigate the line-element of spacetime 
around a linear cosmic string  
in the presence of cosmological constant.
We obtain a static form of the metric and argue how 
it should be discarded because of some asymptotic considerations.
Then a time dependent and consistent form of the metric is obtained
and its properties are discussed. This may be considered as an
example of a preferred frame in physics.
\end{abstract}


\maketitle

\section{\label{sec:intro}Introduction}
 The most reliable measurement of the redshift-magnitude relation
uses supernovae of Type Ia \cite{1}. 
Two groups the High-z Supernova Search Team \cite{2} and the Supernova
Cosmology project \cite{3} working independently and using different
methods of analysis each have found evidence for accelerated
expansion of the universe.
Type Ia supernovae are characterized by 
the absence of hydrogen lines in the spectra and they are thought
to be the result of thermonuclear disruption of white dwarf stars \cite{4}.
The data require $\Lambda >0$ at two or three standard deviations ,
depending on the choice of data and method of analysis \cite{5,6}.
The measurements agree with the relativistic cosmological
model with $\Omega_{k0} =0$ , meaning no space curvature and 
$\Omega_{\Lambda 0}\sim 0.7$ meaning we are living in a cosmological constant dominated universe \cite{7,8,9}. 
Also the latest data of the Wilkinson Microwave Anisotropy Probe are in favour of a flat $\Lambda$-dominated universe\cite{10}. 
These observationally confirmed results and the other
theoretical believes in a non-zero cosmological constant\cite{11} make us to consider the effects
of this parameter on different parts of our studies in cosmology.
In this article we intend to find out its effect on the solution of 
field equations of a cosmic string. Cosmic strings are topologically stable objects which may have formed
during the breaking of a local $U(1)$ gauge symmetry in the very early universe \cite{12,13,14,15}. 
First we find out the static solutions of the Einstein's field equations with non-zero cosmological constant for a straight cosmic string. Then we show that in the limiting case when $\mu\rightarrow 0$ the static form of the solutions do not satisfy the required prediction of the observed slope 5 in the magnitude-redshift relation for low redshifts $(z\leq 0.2)$. Finally time dependent form of the solutions which have consistent asymptotic behaviour are obtained.

\section{\label{sec:static}The static Line-Element}
 In this study we consider an infinitely long, thin, straight , static
string laying along the z-axis with the following stress-energy tensor:
\begin{equation}
T^{\mu}_ {\nu}=\mu \delta(x)\delta(y)diag(1,0,0,1)
\end{equation} \label{1}
\noindent where $\mu$ is the mass per unit length of the string  in the z-direction.

For such a gravitating cosmic string the spacetime possesses the same symmetry and is invariant under time translations, spatial translations in the z-direction, rotation around the z-axis and Lorentz boosts in the z-direction. These special symmetries of the problem, according to the special form of the stress-energy tensor introduced by Eq.(1) guide us to choose the following form of the line-element in the cylindrical coordinate system $(\rho,\phi,z)$ :
\begin{equation}
ds^2=e^{a(\rho)}(dt^2-dz^2)-d\rho^2-e^{b(\rho)}d\phi^2
\end{equation}\label{2}
For the case $\Lambda=0$, the solutions of $a(\rho)$ and $b(\rho)$ are well-known \cite{13,16,17}:
\begin{equation}
ds^2=dt^2-dz^2-d\rho^2-(1-4G\mu)^2\rho^2 d\phi^2
\end{equation}\label{3}
Einstein's field equations with cosmological constant $\Lambda$, in units of $c=1$ , are :
\begin{equation}
R_{\mu \nu} = -8\pi G(T_{\mu \nu}-\frac 12 g_{\mu\nu}T) - \Lambda g_{\mu \nu}
\end{equation} \label{4}
In the case of $\Lambda>0$, $a(\rho)$ and $b(\rho)$ must satisfy the following equations:
\begin{eqnarray}
2b^{^{\prime\prime}}+{b^{^{\prime}}}^2+4a^{^{\prime\prime}}
+2{a^{^{\prime}}}^2&=&-4\Lambda\\
2b^{^{\prime\prime}}+{b^{^{\prime}}}^2+2a^{^{\prime}}b^{^{\prime}}
&=&-4\Lambda\\
2a^{^{\prime\prime}}+2{a^{^{\prime}}}^2+a^{^{\prime}}b^{^{\prime}}
&=&-4\Lambda
\end{eqnarray} 
 \noindent where prime stands for derivation with respect to $\rho$.
These are $\rho\rho$ , $\theta\theta$ and $zz$ components of the field equations for the exterior part of the string, where the stress-energy
tensor and its trace are zero. Actually $tt$ component makes the same equation as the $zz$ component. Combination of Eqs. (5),(6)and (7) yields to the result:
\begin{eqnarray}\label{8,9}
&\;&4a^{^{\prime\prime}}+\frac 34 {a^{^{\prime}}}^2+\Lambda=0\\
&\;&b^{^{\prime}}=2\frac{a^{^{\prime\prime}}}{a^{^{\prime}}}+a^{^{\prime}}
\end{eqnarray}

General solutions of $a(\rho)$ and $b(\rho)$ have the form
\begin{eqnarray}\label{10,11}
a(\rho)&=&\frac 43 \ln{(cos(\frac{\sqrt{3\Lambda}}{2}(\rho+\alpha))}+\beta\\
b(\rho)&=&\frac 43 \ln{(cos(\frac{\sqrt{3\Lambda}}{2}(\rho+\alpha))}
+\ln{(\frac{4\Lambda}{3}tan^2(\frac{\sqrt{3\Lambda}}{2}(\rho+\alpha)))}+\gamma
\end{eqnarray}
where $\alpha , \beta$ and $\gamma$ are constants. To fix these constants
it is sufficient to impose the condition that the metric (2) should match
the form (3) in the limiting case when $\Lambda \rightarrow 0$. With doing this the consistent form of the metric(2) is :
\begin{equation}\label{12}
ds^2=cos^{\frac 43}(\frac{\sqrt{3\Lambda}}{2}\rho)(dt^2-dz^2)-d\rho^2
-\frac{4(1-4G\mu)^2}{3\Lambda}cos^\frac 43(\frac{\sqrt{3\Lambda}}{2}\rho)
tan^2(\frac{\sqrt{3\Lambda}}{2}\rho)d\phi^2
\end{equation}

When $\mu \rightarrow 0$  i.e. in the absence of string , Eq.(12) gives the form of Shwarzschild de Sitter spacetime with $m=0$ in the 
cylindrical coordinates. We have previously shown that in the case of $\Lambda \not = 0$ the so called static isometry of de Sitter solution i.e. 
\begin{equation}\label{13}
ds^2=(1-\frac{\Lambda}{3}\rho^2)dt^2-(1-\frac{\Lambda}{3}\rho^2)^{-1}
d\rho^2-\rho^2(d\theta^2+\sin^2\theta d\phi^2)
\end{equation}
does not fulfill the requirement to predict the observed relation of magnitude redshift for low redshifts $z \leq 0.2$  \cite{18}.
Now let us check this for the metric (12). Evidently in a static spacetime like Eq.(12) the gravitational redshift of a source located at a point with coordinates $(\rho,\theta ,0)$ is proportional to the $00$-component of the metric at that point \cite{19}. In this case calculation of the difference between apparent and absolute magnitudes $m-M$ as a function of redshift $z$ yields:  
\begin{equation}\label{14}
m-M=2.5log[z(2+z)]-2.5log(\frac{\Lambda}{8\pi})
\end{equation}
Inspecting the logarithmic slope of Eq.(14) for small values of $z$ , it turns out to be $2.5$. Then a comparison of Eq.(14) with the so called redshift-magnitude relation in a FRW model reveals a shortcoming of the static metric (12) in predicting the experimentally tested value of slope $5$ \cite{18}.
So it provides enough motivation to seek for a nonstatic solution that overcomes this deficiency. Now we continue to find  non-static solutions of the field equations for cosmic strings.

\section{\label{sec:nonstatic} The Non-Static Line-Element}
To investigate the nonstatic solution of the cosmic strings we may choose the metric Eq.(2) multiplied by a scaling factor, i.e.
\begin{equation}\label{15}
ds^2={\cal S}^2(t) \left\{ e^{a(\rho)}(dt^2-dz^2)-d\rho^2-e^{b(\rho)}d\phi^2
\right\}
\end{equation}
To find the unknown functions of the metric
it reveals to be more simple if we 
rescale the time coordinate to write the line-element in the form
\begin{equation}\label{16}
ds^2=e^{a(\rho)}d\tau^2-{\cal R}^2(\tau)\left\{d\rho^2+e^{b(\rho)}d\phi^2+e^{a(\rho)}dz^2\right\}
\end{equation}
It remains to solve the field equations to determine the functions $a$ , $b$ and ${\cal R}$. Direct calculations of the Ricci tensor lead to the following field equations :
\begin{eqnarray}\label{17,18,19,20,21}
2a^{^{\prime\prime}}+2{a^{^{\prime}}}^2-12{\ddot{\cal R}}{\cal R} e^{-a}+
a^{^{\prime}}b^{^{\prime}}&=&-4\Lambda {\cal R}^2\\
-4\ddot{\cal R}{\cal R}e^{-a}+4a^{^{\prime\prime}}+2{a^{^{\prime}}}^2
+2b^{^{\prime\prime}}-8{\dot{\cal R}}^2e^{-a}+{b^{^{\prime}}}^2
&=&-4\Lambda{\cal R}^2\\
-4\ddot{\cal R}{\cal R}e^{-a}+2a^{^{\prime}}b^{^{\prime}}
+2b^{^{\prime\prime}}+{b^{^{\prime}}}^2-8{\dot{\cal R}}^2e^{-a}
&=&-4\Lambda{\cal R}^2\\
-4\ddot{\cal R}{\cal R}e^{-a}+2{a^{^{\prime}}}^2+2a^{^{\prime\prime}}
-8{\dot{\cal R}}^2e^{-a}+a^{^{\prime}}b^{^{\prime}}&=&-4\Lambda{\cal R}^2\\
\dot{\cal R}a^{^{\prime}}&=&0 
\end{eqnarray} 
where prime and dot indicate differentiation with respect to $\rho$
and $\tau$ respectively. A physically nontrivial solution of Eq.(21) may be
treated as $a$ to be constant. To be consistent with Eq.(3) when $\Lambda \rightarrow 0$ this constant should be equal to zero. In this case the Eq.(17) yields the following result for $\cal R$;
\begin{equation}\label{22}
3\ddot{\cal R}{\cal R}={\cal R}^2\Lambda
\end{equation}
So that
\begin{equation}\label{23}
{\cal R}=e^{\sqrt{\frac{\lambda}{3}}\tau}
\end{equation}
This in turn results that:
\begin{equation}\label{24}
2b^{^{\prime\prime}}+{b^{^\prime}}^2=0
\end{equation}
Evidently the solution $b=const.$ satisfies the Eq.(24).
If we accept that the metric (3) should be recovered in the 
limiting case $\Lambda\rightarrow 0$, consequently this constant should be 
equal to $(1-4G\mu)^2$. Therefore the nonstatic line-element of the cosmic string is:
\begin{equation}\label{25}
ds^2=dt^2-e^{^{2\sqrt{\frac{\Lambda}{3}}t}}\left\{ d\rho^2+(1-4G\mu)^2d\phi^2
+dz^2\right\}
\end{equation}

By a transformation of polar angle, $\theta\rightarrow(1-4G\mu)\theta$,
the metric becomes the flat-space deSitter metric. As expected , spacetime around a cosmic string is that of empty space. However, the range 
of the flat-space polar angle $\theta$ is only $0\leq\theta\leq 2\pi(1-4G\mu)$
rather than $0\leq\theta\leq 2\pi$. Due to this and nonstatic nature
of (25), the observer will see two images of the source , with the 
angle seperation $\delta\alpha$ between the two images defined by
\begin{equation}\label{26}
\delta\alpha=8\pi G\mu \frac{l}{l+d} (1-\sqrt{\frac{\Lambda}3}l)
\end{equation}
Here $l$($d$) is the distance from string to the source (observer)
at the observation epoch. It is valid for the approximation range of $\sqrt{\frac{\Lambda}3}(l+d)$ to be small compared to 1. Thus the existence of cosmological constant will cause to weaken the double images of objects
located behind the string. The other important point to mention 
concerning (25) is that like deSitter spacetime ,it possesses
an event horizen at the distance $(\sqrt{\frac{3}{\Lambda}})$.
We conclude that in the presence of cosmological constant this 
special frame is preferred by the cosmological considerations.


\end{document}